# Beryllium Silicide Clusters, $Be_nSi_n$, $Be_{2n}Si_n$ (n = 1 – 4) and possible $MgB_2$-like Superconductivity in some of them

O. P. Isikaku-Ironkwe[1, 2]
[1]The Center for Superconductivity Technologies (TCST)
Department of Physics,
Michael Okpara University of Agriculture, Umudike (MOUAU),
Umuahia, Abia State, Nigeria
and
[2]RTS Technologies, San Diego, CA 92122

## Abstract

The symmetry of the Periodic Table makes it possible to predict certain properties of similar elements and compounds using one of them as a template. Magnesium diboride, $MgB_2$, presents a useful template in the search for similar materials. Starting from electronegativity, valence electron and atomic number equivalency, we identify many potential similar materials. One of them is the beryllium silicide, $Be_{2n}Si_n$ cluster system. We establish that though not yet produced in bulk, $Be_2Si$ exists. We show from symmetry principles that beryllium silicide and some of its clusters will be $MgB_2$-like superconductors with Tcs close to or higher than 39K.

## Introduction

The discovery of superconductivity in $MgB_2$ in 2001 [1] came as a surprise to many researchers in superconductivity. A bigger surprise also came when application of electronic structure principles [2 - 5] and crystal symmetry concepts [6 – 13] failed to produce $MgB_2$-like superconductors with Tcs close to the 39K of $MgB_2$. In searching for $MgB_2$-like superconductors and other superconductors, we decided to look for materials specific correlations, often ignored by standard theories of superconductivity. Some of the chemical correlations with superconductivity are electronegativity, valence electron count, atomic number and formula weight. Our studies [14 - 18] indicate that they play a very significant role in the search for superconductivity. In this paper, we first review studies [19 – 24] on the existence, properties and preparation of beryllium silicide clusters. We then use the material specific characterization system (MSCD), symmetry rules [15], and the Tc equation to compare the material specific characteristics of $Be_{2n}Si_n$ clusters which have same valence electron count and atomic number as MgB$_2$ and to estimate the Tc of Be$_2$Si, based on this symmetry.

**Beryllium Silicide Studies**
Group IIA-IV alkaline-earth silicides have been studied by the *ab initio* pseudopotential method [19] and the full-potential linearized augmented plane wave method [20], within the local density approximation (LDA). This family is found to have an anti-fluorite structure and semiconducting. Beryllium silicide, Be$_2$Si, a member of this family was however computed to be a metallic [19], antifluorite, with Si atoms in face-centered cubic structure and the Be atoms arranged around them in tetrahedral structure. In addition, the Be-Si bonding has been studied [21, 22] by density functional theory (DFT) Monte Carlo simulated annealing (DFT-MCSA) methods and cluster geometries discovered. In particular, the structures and energetics of Be$_n$Si$_n$ and Be$_{2n}$Si$_n$ (n = 1 – 4) clusters [21, 22] strongly suggest that they could be explored for superconductivity like the carbon-60 fullerene [23]. Hite et al [24] have explored the possible formation of Beryllium silicide. Using scanning tunneling microscopy (STM) and photoelectron spectroscopy, they studied the nucleation, growth and structure of beryllium on Si(111)-(7x7)surface at temperatures ranging from 120K to 1175K. They produced amorphous ring nanocluster structure with Be-Si bond length less than 2.5 A of Beryllium silicide. Further studies by Saranin et al. [25] confirmed the cluster structure and discovered four types of nanostructure arrays formed by Be interaction with Si(111)-(7x7) surface in a not fully understood "self assembly" process.

**MSCD of Be$_2$Si, Symmetry Rules and Tc Estimation**
In [15, 16] we showed that a material may be characterized in terms of averages of electronegativity, $\mathcal{X}$, valence electron count, Ne, atomic number, Z, and formula weight, Fw, in a method described as material specific characterization dataset (MSCD). MSCD makes it possible to quickly compare two or more materials. Table 1 gives the MSCD of MgB$_2$, LiBSi and Be$_{2n}$Si$_n$ clusters. We also showed in [15] that the maximum transition temperature, T$_c$, of a superconductor can be estimated in material specific properties of electronegativity, $\mathcal{X}$, valence electrons, Ne, atomic number, Z, and a parameter, Ko by the equation:

$$T_c = x \frac{Ne}{\sqrt{Z}} K_o \tag{1}$$

where $K_o$ is a parameter that determines the value of Tc. $K_o$ = n(Fw/Z) and n is dependent on the family of superconductors. Fw represents formula weight of the superconductor. For $MgB_2$, with Tc of 39K and Fw/Z of 6.26, Ko = 22.85, making n = 3.65. Note that the MSCDs of LiBSi and $Be_2Si$ are identical. One of the symmetry rules [15] states that two materials with exactly identical MSCDs will have the same Tc. Following this symmetry rule and the MSCD displayed in Table 1, we estimate that the Tc of $Be_2Si$ will be the same as that of LiBSi [17], which is 36K.

## Discussion

Even though bulk $Be_2Si$ has yet to be produced and fully characterized, recent computations [19, 21, 22] and experiments [24, 25] strongly suggest that beryllium silicide exists. $Be_2Si$ is an image of LiBSi, formed by replacing LiB with 2 atoms of Be. Thus $Be_2Si$ and LiBSi have the same electronegativity, valence electrons and atomic number. $Be_2Si$ also has the same valence electron count and atomic number as $MgB_2$ making it an $MgB_2$-like material. From the symmetry rules for searching for superconductors [15], materials with the same valence electron count and same atomic number will have Tcs proportional to their electronegativities. The challenge of producing bulk $Be_2Si$ is akin to that of producing bulk $C_{60}$ fullerene [23, 26]. We note too that well-ordered binary cluster $Cs_3C_{60}$ attained a Tc of 38K [27]. It will be interesting if well-ordered $Be_2Si$ could achieve a higher Tc than the 38K of $Cs_3C_{60}$. The verification (or otherwise) of superconductivity in $Be_2Si$ and some of its clusters will be a strong test of the symmetry rules [15, 18] for searching for superconductors.

## Conclusion

We have studied the hypothetical cluster compound beryllium silicide. Using symmetry rules for searching for equivalent superconductors and $MgB_2$ as a template, we find that $Be_2Si$, $Be_4Si_2$, $Be_8Si_4$ are copies of $MgB_2$ but with reduced electronegativity. $Be_2Si$ too is very similar to LiBSi, which we had earlier shown [17] should be superconducting. We conclude that $Be_2Si$ will be found to be superconducting with a Tc of about 36.2K. The clusters: $Be_4Si_2$, $Be_6Si_3$, $Be_8Si_4$ may have Tcs higher than $MgB_2$ since they have higher Fw/Z ratio [15].

## Acknowledgements

The author acknowledges financial support for this research from M.J. Schaffer, then at General Atomics, enlightening discussions with M.B. Maple at UC San Diego and literature support from J.R. O'Brien at Quantum Design, San Diego.

## TABLE

Table 1: MSCDs of MgB2, LiBSi and Beryllium Silicide Clusters $Be_{2n}Si_n$ (n = 1 – 4) computed from equations in ref. [15] and equation (1) in this paper. Note that LiBSi and Be2Si have exactly the same electronegativity, , valence electron count, Ne, atomic number Z and almost the same formula weight Fw. Symmetry rule given in [15] predicts that their Tcs will be about the same.

|   | Material | $x$ | Ne | Z | $Ne/\sqrt{Z}$ | Fw | Fw/Z | Tc(K) | Ko |
|---|---|---|---|---|---|---|---|---|---|
| 1 | $MgB_2$ | 1.7333 | 2.6667 | 7.3333 | 0.9847 | 45.93 | 6.263 | 39 | 22.85 |
| 2 | LiBSi | 1.6 | 2.6667 | 7.3333 | 0.9847 | 45.84 | 6.251 | 35.95 | 22.85 |
| 3 | $Be_2Si$ | 1.6 | 2.6667 | 7.3333 | 0.9847 | 46.11 | 6.288 | 36.2 | 22.85 |
| 4 | $Be_4S_2$ | 1.6 | 2.6667 | 7.3333 | 0.9847 | 92.22 | 12.58 | >39? | 22.85? |
| 5 | $Be_6S_3$ | 1.6 | 2.6667 | 7.3333 | 0.9847 | 138.33 | 18.86 | >39? | 22.85? |
| 6 | $Be_8S_4$ | 1.6 | 2.6667 | 7.3333 | 0.9847 | 184.44 | 25.15 | >39? | 22.85? |